\documentclass{webofc}
\usepackage[varg]{txfonts}   
\begin{document}
\title{Heavy-quark effects on cold quark matter and self-bound stars}
%
%

\author{\firstname{José C.} \lastname{Jiménez}\inst{1}\fnsep\thanks{\email{jimenez@if.usp.br}} \and
        \firstname{Eduardo} \lastname{S. Fraga}\inst{2}\fnsep\thanks{\email{fraga@if.ufrj.br}}}

\institute{Instituto de F\'isica, Universidade de S\~ao Paulo, \\ 
Rua do Mat\~ao, 1371, Butant\~a,  05508-090, S\~ao Paulo, SP, Brazil
\and
Instituto de F\'\i sica, Universidade Federal do Rio de Janeiro, \\
Caixa Postal 68528, 21941-972, Rio de Janeiro, RJ, Brazil 
}

\abstract{%
 Heavy-quark effects on the equation of state for cold and dense quark matter are obtained from perturbative QCD, yielding observables parametrized only by the renormalization scale. In particular, we investigate the thermodynamics of charm quark matter under the constraints of $\beta$ equilibrium and electric charge neutrality in a region of densities where perturbative QCD is, in principle, much more reliable. Finally, we analyze the stability of charm stars, a possible new branch of ultradense, self-bound compact stars, and find that they are unstable under radial oscillations.
}
\maketitle
\section{Introduction}
\label{intro}

The presence of heavy flavors in cold quark matter might have non-trivial effects on matter subject to extreme conditions of density. For instance, the case of charm quarks immediately leads to the possibility of charm stars, i.e. self-bound strange stars with some amount of charm quarks in their cores. However, in spite of some constraints on their existence \cite{Lattimer:2010uk,Lattimer:2019eez}, they might still occur in ultradense hybrid neutron stars. 
Perturbative quantum chromodynamics (pQCD) can be used much more confidently at these very high densities \cite{Kapusta:2006pm} to build the equation of state (EoS) for charm matter\footnote{This complements advancements in pQCD \cite{Laine:2006cp} and lattice QCD at high temperatures \cite{Borsanyi:2016ksw}.} in electroweak equilibrium. 

We investigate cold quark matter with heavy quarks using in-medium pQCD. Specifically, we extend the framework developed in Ref. \cite{Kurkela:2009gj} for $N_{f}=N_{l}+1$ flavors, i.e. $N_{l}$ massless quarks and $1$ massive flavor, to include any number of massive flavors and determine the associated EoS. As an application, we discuss charm matter that is in $\beta$ equilibrium and electrically neutral. This allows us to consider the existence charm quark stars from the pQCD viewpoint. In the past, these exotic stars were investigated only within the bag model, being unstable against radial oscillations \cite{Kettner:1994zs}. Using the method developed in Ref. \cite{Gondek:1997fd} for a couple of first-order differential equations for the Lagrangian radial displacements, our calculations indicate that such self-bound stars are unstable.

\section{Cold and dense pQCD with heavy quarks}
\label{sec1:framework}

The calculation of the EoS for cold quark matter within pQCD has a long history \cite{Kapusta:2006pm}. The three-loop result of Ref. \cite{Kurkela:2009gj} includes renormalization-group effects on the mass and strong coupling parameters. As mentioned above, this framework deals with $N_{f}=N_{l}+1$ quark flavors. The corresponding  thermodynamic potential has a massless contribution plus a massive term, together with the mixed vacuum-matter (VM) diagrams and the corresponding ring terms, i.e. $\Omega=\Omega^{m=0}(\vec{\mu})+\Omega^{m}(\tilde{\mu},m)+\Omega^{x}_{{\rm VM}}(\vec{\mu}, m)+\Omega_{{\rm ring}}(\vec{\mu}, \tilde{\mu}, m)$. Here $m$ is the mass, $\tilde{\mu}$ corresponds to the massive flavor chemical potential and $\vec{\mu}~{\equiv}~(\mu_{1}, ...,\mu_{N_{l}})$ is the vector chemical potential for the massless quarks. $\Omega$ also depends on the renormalization scale parameter $\bar{\Lambda}$ and  $\alpha_{s}(\bar{\Lambda})=({4\pi}/{\beta_{0}L})\left(1-({2\beta_{1}}/{\beta^{2}_{0}}){\ln{L}}/{L}\right)$, where $\beta_{0}=11-2N_{f}/3$, $\beta_{1}=51-19N_{f}/3$, $L=2\ln\left(\bar{\Lambda}/\Lambda_{\rm \overline{MS}}\right)$ and $\Lambda_{\rm \overline{MS}}$ is the $\overline{\rm MS}$ point scale. Usually the renormalization scale is defined as $\bar{\Lambda}=X\sum_i{\mu_{i}}/N_{f}$, where the sum runs over the active quark flavors and the dimensionless parameter $X$ sits between $1$ and $4$ \cite{Kurkela:2009gj}. 

On the other hand, at high densities one expects that light quarks could be accompanied by heavy flavors. In this situation, the total number of flavors is $N_{f}=\sum^{N_{m}}_{i=1}(N_{l}+1)^{(i)}$, where $N_{m}$ is the number of massive quarks. This restricts us to add at least one massless quark for each massive flavor. Notice that further physical conditions are required in order to control when a heavy partner is active. Such conditions affect the renormalization scale $\bar{\Lambda}$, depending on the chosen heavy quark to be added. Thus, $\Omega$ at $\mathcal{O}(\alpha^{2}_{s})$ for $N_{l}$ massless and $N_{m}$ massive quarks is $\bar{\Omega}[{N_{f}}]= \sum^{N_{m}}_{i=1}\left\lbrace\Omega[{N^{(i)}_{l}}]+\Omega[{1^{(i)}}]\right\rbrace$, choosing first the massless flavors when adding a massive one, so that $\Omega[{N^{(i)}_{l}}]~{\equiv}~(\Omega^{m=0}(\vec{\mu}))^{(i)}$ is the massless contribution and $\Omega[{1^{(i)}}]~{\equiv}~(\Omega^{m}+\Omega^{x}_{{\rm VM}}+\Omega_{{\rm ring}})^{(i)}$ is the mixed massive contribution. Here $\vec{\mu}^{(i)}=(\mu_{1}, ...,\mu_{i})$ is the massless vector chemical potential, $\tilde{\mu}^{(i)}$ the massive (heavy) quark chemical potentials and $m^{(i)}$ their corresponding masses.  $\Omega[...]$ indicates just the implicit parameter dependence (e.g. on $N_{f}$), whereas $\Omega(...)$ represents an explicit function dependence.

\section{Charm matter thermodynamics}
\label{sec2:charm}

We now apply this formalism to charm quark matter. Below the charm threshold, the usual strange matter conditions apply \cite{Glendenning:2000}. Above the threshold, the reactions ${u+d}\leftrightarrow{c+d}$ yield $\mu_{c}=\mu_{u}$. Furthermore, electric neutrality implies $({2}/{3})n_{u}+({2}/{3})n_{c}-({1}/{3})n_{d}-({1}/{3})n_{s}-n_{e}-n_{\mu}=0$, with $\mu_{\mu}=\mu_{e}$, so that all chemical potentials are functions of $\mu_{s}$. The strange and charm quark mass parameters (up to $\mathcal{O}(\alpha^{2}_{s})$) are \cite{Vermaseren:1997fq} $m_{s}(\bar{\Lambda})=\hat{m}_{s}\left({\alpha_{s}}/{\pi}\right)^{4/9}\left(1+0.895062\left({\alpha_{s}}/{\pi}\right) +1.37143\left({\alpha_{s}}/{\pi}\right)^{2}\right)$ and $m_{c}(\bar{\Lambda})=\hat{m}_{c}\left({\alpha_{s}}/{\pi}\right)^{12/25}\left(1+1.01413\left({\alpha_{s}}/{\pi}\right)+1.38921\left({\alpha_{s}}/{\pi}\right)^{2}\right)$, respectively, $\lbrace\hat{m}_{q}\rbrace$ being the invariant quark masses. By noting that $\alpha_{s}$ depends on $N_{f}$, this gives different values of $\Lambda_{\overline{\rm MS}}$. Choosing $\alpha_{s}(\bar{\Lambda}=1.5~{\rm GeV},~N_{f}=3,4)=0.336^{+0.012}_{-0.008}$ \cite{Bazavov:2014soa} gives $\Lambda^{2+1}_{\overline{\rm MS}}=343^{+18}_{-12}~$MeV and $\Lambda^{2+1+1}_{\overline{\rm MS}}=290^{+18}_{-12}~$MeV, thus defining $\alpha^{2+1}_{s}(\bar{\Lambda})$ and $\alpha^{2+1+1}_{s}(\bar{\Lambda})$, respectively. Fixing the strange quark mass at $m_{s}(2~{\rm GeV}, N_{f}=3,4)=92.4(1.5)~$MeV \cite{Chakraborty:2014aca} gives $\hat{m}^{2+1}_{s}~{\approx}~246.2~$MeV when using $\alpha^{2+1}_{s}$, and $\hat{m}^{2+1+1}_{s}~{\approx}~243.7~$MeV with $\alpha^{2+1+1}_{s}$. Additionally, fixing the charm quark mass at $m_{c}(3~ {\rm GeV}, N_{f}=4)=0.9851(63)~{\rm GeV}{~\equiv~}m^{0}_{c}$ \cite{Chakraborty:2014aca}, gives $\hat{m}^{2+1+1}_{c}~{\approx}~3.0895~$GeV when using $\alpha^{2+1+1}_{s}$, being $m^{0}_{c}$ the vacuum charm mass. Furthermore, charm quarks must satisfy $\mu_{c}=\mu_{s}-\mu_{e}>m^{\rm medium}_{c}>m^{0}_{c}$ being $m^{\rm medium}_{c}$ the unknown in-medium charm mass. The renormalization scale parameter below and above the charm threshold is ${\bar{\Lambda}=}X{(\mu_{u}+\mu_{d}+\mu_{s}+0)}/{3}$ for $\mu_{s}~{\lesssim}~m^{0}_{c}$ and $X^{*}{(\mu_{u}+\mu_{d}+\mu_{s}+\mu_{c})}/{3}$ for $\mu_{s}~{\gtrsim}~m^{0}_{c}$, 
where the approximations in the inequalities come from noticing that before the threshold point the electron chemical potential takes its lowest value compared to the strange one, thus allowing us to make the approximation $\mu_{c}~{\approx}~\mu_{s}$.
\begin{figure*}[t!]
\hbox{
\includegraphics[width=0.5\textwidth]{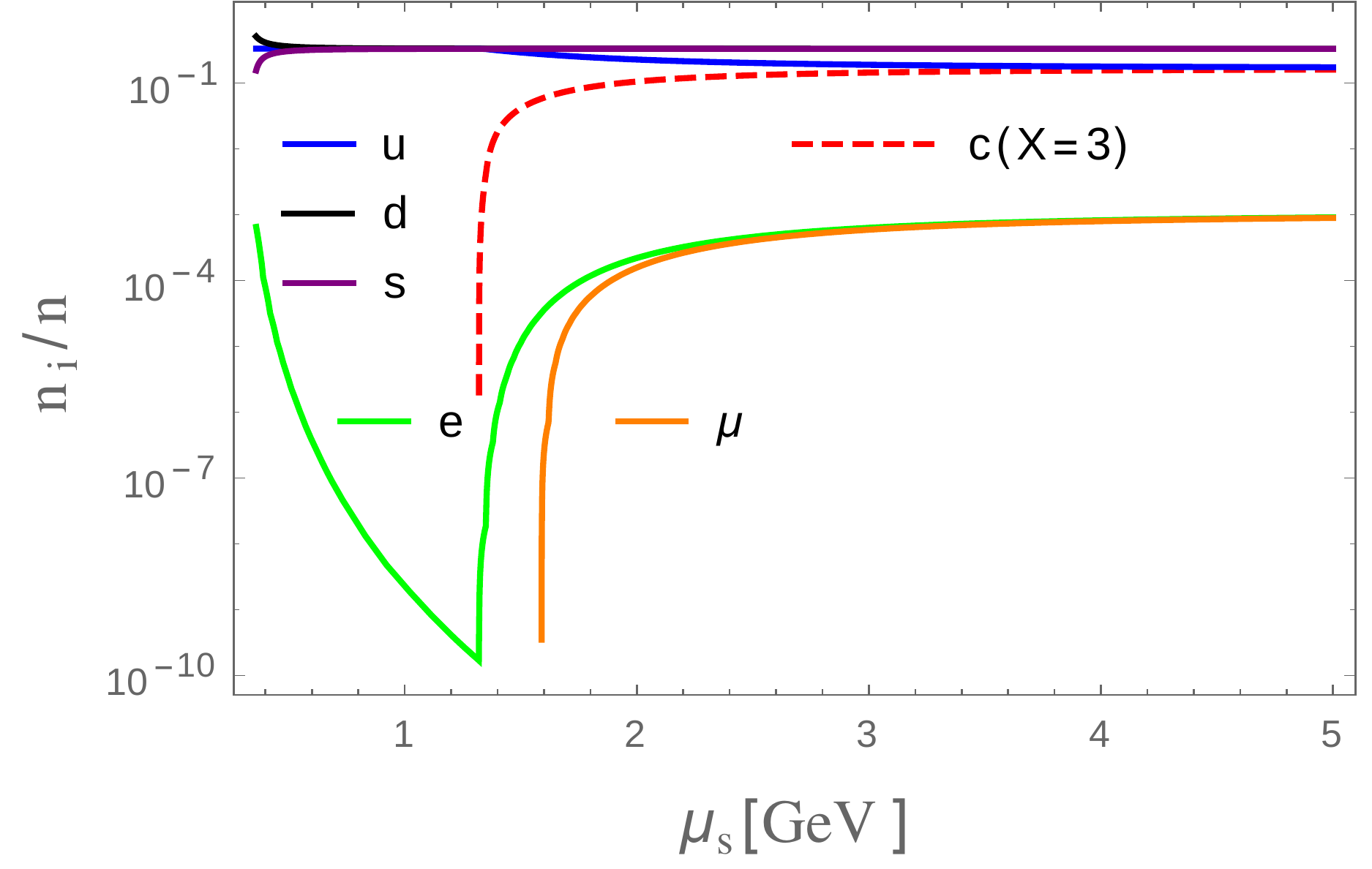} 	\includegraphics[width=0.5\textwidth]{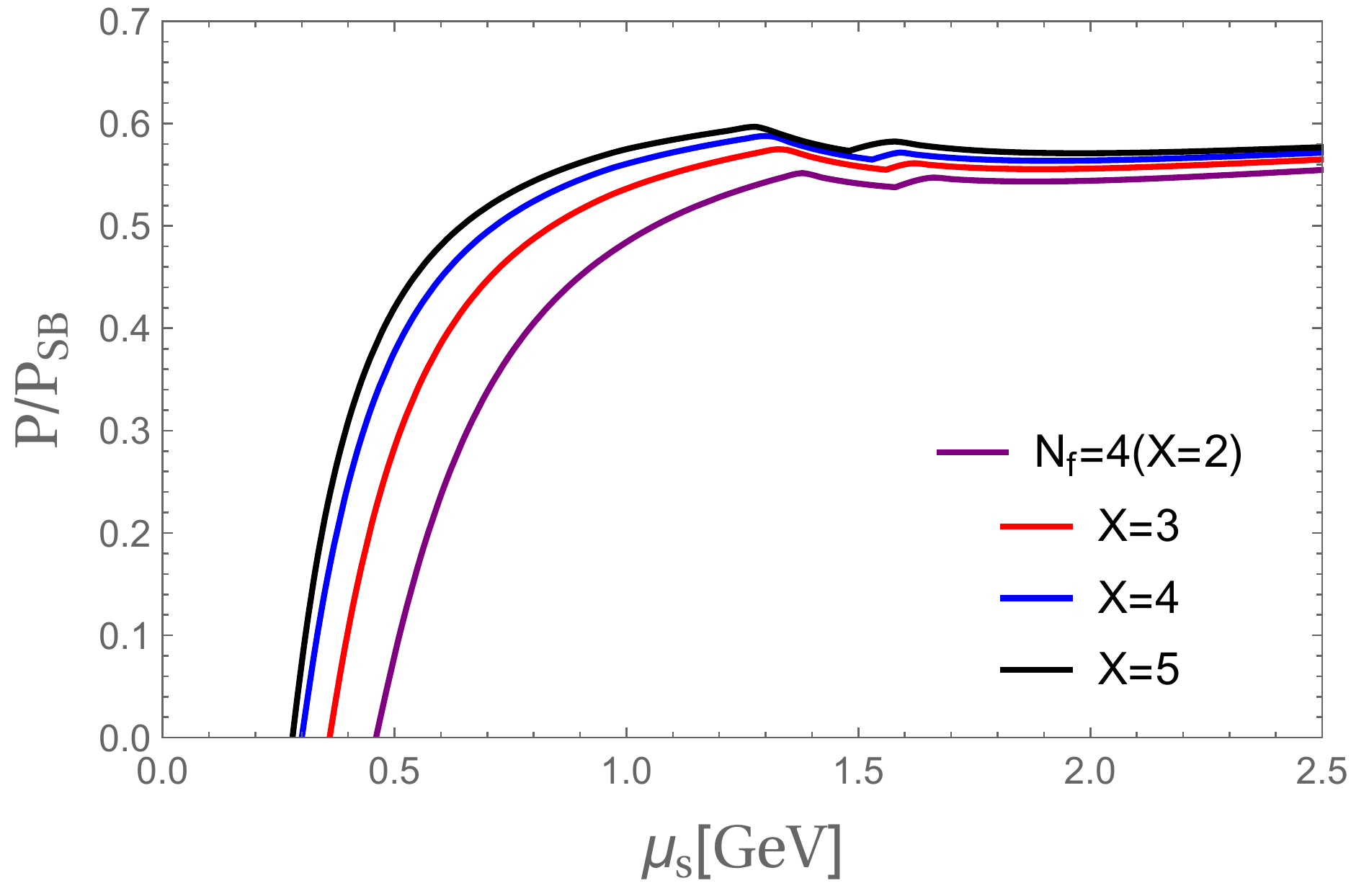}}
\caption{Left panel: Quark and lepton populations, $n_{i}/n$, present in charm matter for $X=3$. Above the charm threshold, the lepton fractions increase to ensure electric charge neutrality at high densities. Right panel: Total pressure of $N_{f}=2+1+1$ quarks plus leptons in $\beta$ equilibrium and electrically charge neutral normalized by the Stefan-Boltzmann gas with $N_{f}=4$. The case $X=5$ is shown only to verify how the EoS depends on $X$ when including one additional heavy flavor. Taken from Ref. \cite{Jimenez:2019kji}.}
\label{fig:thermo}
\end{figure*}
To build the EoS, we choose $N^{(1)}_{l}=1$ for the up, $N^{(2)}_{l}=1$ for the down, and $N_{m}=2$ for the strange and charm quarks in the thermodynamic potential $\bar{\Omega}[N_{f}=2+1+1]=\left\lbrace\Omega[N^{(1)}_{l}=1]+\Omega[1^{(1)}]\right\rbrace+\left\lbrace\Omega[N^{(2)}_{l}=1]+\Omega[1^{(2)}]\right\rbrace$, so that the flavors are counted as $N_{f}=(1+1)^{(1)}+(1+1)^{(2)}=(u+c)^{(1)}+(d+s)^{(2)}$. We define the total quark number density for charm matter, for a given $X$, as $n_{q}~{\equiv}~n_{u}+n_{d}+n_{s}+n_{c}$ and the total particle density as $n=n_{q}+n_{L}$, where $n_{L}=n_{e}+n_{\mu}$. In Fig. \ref{fig:thermo}, we show the behavior of the relative particle populations in $\beta$ equilibrium and electrically neutral charm quark matter in the case of $X=3$. The total pressure and energy density are obtained by keeping thermodynamic consistency systematically. This allows us to build numerically the EoS, $P=P(\epsilon, X)$. In Fig. \ref{fig:thermo}, the total pressure is displayed for charm matter normalized by a Stefan-Boltzmann gas of quarks with $N_{f}=4$ as a function of the strange quark chemical potential. In the same figure, we show that the charm EoS is largely softened, generating an apparent instability which could have astrophysical effects, e.g. the possibility of charm stars.
\begin{figure*}[t!]
\hbox{
\includegraphics[width=0.5\textwidth]{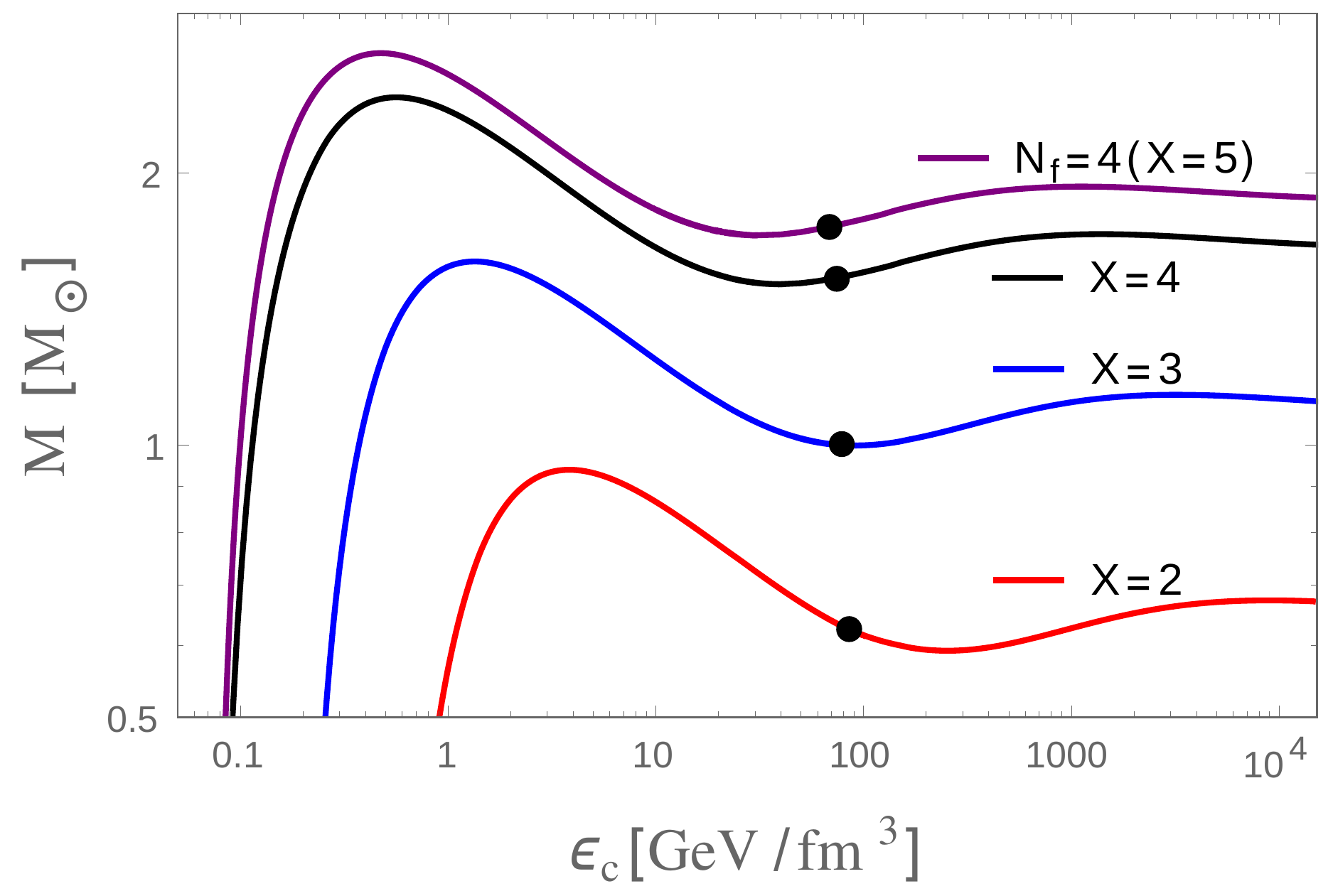} 	\includegraphics[width=0.5\textwidth]{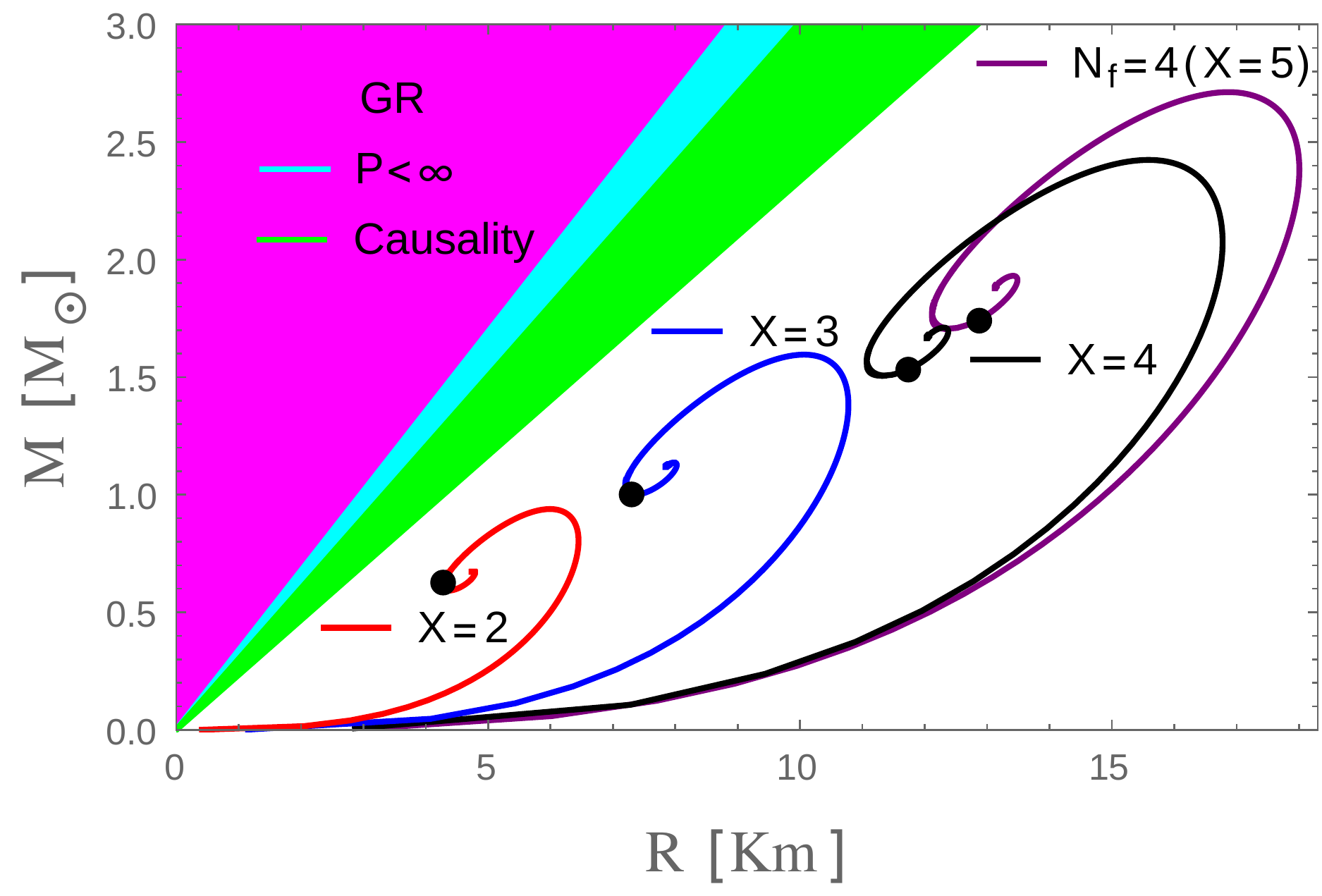}}
\caption{Left panel: Gravitational masses vs central energy densities for a system with $N_{f}=2+1+1$ quarks plus leptons. Right panel: Mass-radius diagram for the associated quark stars. The black dots indicate the appearance of the charm quarks. Stars not satisfying the general-relativistic (GR, magenta), causality (green), and finite pressure ($P<\infty$, light blue) limits are excluded \cite{Lattimer:2006xb}. Taken from Ref. \cite{Jimenez:2019kji}.}
\label{fig:diagrams}
\end{figure*}
%
\section{Results and Conclusion}
\label{sec3:results}

The possibility of bare charm quark stars was studied in Ref. \cite{Kettner:1994zs} using the simplest version of the bag model. Their conclusion was that charm stars would be unstable against radial perturbations. We revisit this question using our first-principle pQCD results for the EoS for charm quark matter. We also choose the range $X\geq3$, which satisfies the Bodmer-Witten hypothesis of strange quark matter. Although the structure equations ensure the relativistic hydrostatic equilibrium of the stars, they must also satisfy the condition $\partial{M}/\partial{\epsilon_{c}}~{\geq}~0$ \cite{Glendenning:2000}. The maximal masses occur at $\partial{M}/\partial{\epsilon_{c}}=0$ (see Fig. \ref{fig:diagrams}). For the case $X=2$, this condition is not satisfied when charm quarks appear. On the other hand, for $X>3$, this condition is satisfied when charm quarks are present, which would correspond to stable charm stars. In Fig. \ref{fig:diagrams}, we show the mass-radius diagram for quark stars made of $N_{f}=2+1+1$ quarks plus electrons and muons with varying $X$. Performing the dynamical stability analysis against radial pulsations using the method of Ref. \cite{Gondek:1997fd}, amplitudes oscillate harmonically when the frequencies are such that ${\rm Re(\omega_{n})>0}$ and ${\rm Im(\omega_{n})=0}$, or increase exponentially if ${\rm Re(\omega_{n})~{\geq}~0}$ and ${\rm Im(\omega_{n})>0}$.  Besides, if ${\rm Im(\omega_{0})>0}$ [and ${\rm Re(\omega_{0})=0}$] from some value of central energy density $\epsilon_{c}$, then all the higher modes will become complex too, representing the onset of the instability. Our calculations show that for $X=3$ the amplitudes of the stellar configurations increase even in the region where charm stars would be expected, making them dynamically unstable. Since the same behavior was obtained for larger values of $X$, one can conclude from a perturbative QCD analysis that charm stars are unstable. In the future, it would be interesting to study the presence of charm quarks in neutron-star mergers.

\section*{Acknowledgement}

This work was partially supported by INCT-FNA (Process No. 464898/2014-5). J. C. J. acknowledges support from FAPESP (Processes No. 2020/07791-7 and No. 2018/24720-6). E. S. F. is partially supported by CAPES (Finance Code 001), CNPq, and FAPERJ.


\end{document}